\documentclass[aps,preprint,showpacs,groupedaddress,floatfix]{revtex4}
\usepackage{graphicx}
\usepackage{dcolumn}
\usepackage{bm}

\begin{document}

\title{Renormalized relativistic Hartree-Bogoliubov 
equations with a zero-range pairing interaction}
\author{T. Nik\v si\' c}
\author{P. Ring}
\affiliation{Physik-Department der Technischen Universit\"at M\"unchen, 
D-85748 Garching,
Germany}
\author{D. Vretenar}
\affiliation{Physics Department, Faculty of Science, University of Zagreb, 
Zagreb, Croatia}
\date{\today}

\begin{abstract}
A recently introduced scheme for the renormalization of the 
Hartree-Fock-Bogoliubov equations in the case of zero-range 
pairing interaction is extended to the relativistic 
Hartree-Bogoliubov model. 
A density-dependent strength parameter of the zero-range pairing is adjusted
in such a way that the renormalization procedure reproduces the empirical 
$^1S_0$ pairing gap in isospin-symmetric nuclear matter. 
The model is applied  
to the calculation of ground-state pairing properties of 
finite spherical nuclei.
\end{abstract}

\maketitle 

\section{Introduction}

The theoretical framework of self-consistent 
mean-field models enables a description of the nuclear 
many-body problem in terms of universal energy density functionals. 
By employing global effective interactions,
adjusted to empirical properties of symmetric and asymmetric 
nuclear matter, and to bulk properties of 
spherical nuclei, the current generation of 
self-consistent mean-field models has achieved a high level of 
accuracy in the description of ground states and
properties of excited states in arbitrarily heavy nuclei,
exotic nuclei far from $\beta$-stability,
and in nuclear systems at the nucleon drip-lines \cite{BHR.03}.

The relativistic mean-field (RMF) models, 
in particular, are based on concepts of non-renormalizable effective
relativistic field theories and density functional theory.
They have been very successfully applied in studies of nuclear
structure phenomena at and far from the valley of $\beta$-stability.
For a quantitative analysis of open-shell   
nuclei it is necessary to consider also pairing correlations.
Pairing has often been taken into account in a very
phenomenological way in the BCS model with the monopole pairing
force, adjusted to the experimental odd-even mass differences. This
approach, however, presents only a poor approximation for
nuclei far from stability. The physics of
weakly-bound nuclei necessitates a unified and self-consistent
treatment of mean-field and pairing correlations. This has led to
the formulation and development of the relativistic
Hartree-Bogoliubov (RHB) model, which represents a relativistic
extension of the conventional Hartree-Fock-Bogoliubov framework,
and provides a basis for a consistent microscopic 
description of ground-state properties of  
medium-heavy and heavy nuclei, low-energy excited states, 
small-amplitude vibrations, and 
reliable extrapolations toward the drip lines \cite{VALR.05}.

In most applications of the RHB model \cite{VALR.05}
the pairing part of the well 
known and very successful Gogny force~\cite{BGG.84} has be employed in the
particle-particle ($pp$) channel:
\begin{equation}
V^{pp}(1,2)~=~\sum_{i=1,2}e^{-((\mathbf{r}_{1}-\mathbf{r}_{2})/{\mu_{i}})^{2}%
}\,(W _{i}~+~B_{i}P^{\sigma}-H_{i}P^{\tau}-M_{i}P^{\sigma}P^{\tau})\;,
\end{equation}
with the set D1S \cite{BGG.91} for the parameters $\mu_{i}$,
$W_{i}$, $B_{i}$, $H_{i}$, and $M_{i}$ $(i=1,2)$. This force has
been very carefully adjusted to the pairing properties of finite
nuclei all over the periodic table. In particular, the basic
advantage of the Gogny force is the finite range, which
automatically guarantees a proper cut-off in momentum space.
However, the resulting pairing field is non-local and the solution of the 
corresponding Dirac-Hartree-Bogoliubov integro-differential equations
can be time-consuming, especially in the case of deformed nuclei.
Another possibility is the use of a zero-range, possibly density-dependent,
$\delta$-force in the $pp$-channel of the RHB model \cite{Meng.98}. This
choice, however, introduces an additional cut-off parameter in
energy and neither this parameter, nor the strength of the
interaction, can be determined in a unique way. The effective range of 
the interaction is determined by the energy cut-off, and the 
strength parameter must be chosen accordingly in order to reproduce
empirical pairing gaps. 

In a series of recent papers \cite{Bul1,Bul2,Bul3} A. Bulgac and Y. Yu
have introduced a simple scheme for the renormalization of the 
Hartree-Fock-Bogoliubov equations in the case of zero-range 
pairing interaction. The scheme is equivalent to a simple energy cut-off
with a position dependent coupling constant. In this work we use 
the prescription of Refs.~\cite{Bul1,Bul2} to implement a 
regularization scheme for the relativistic Hartree-Bogoliubov 
equations with zero-range pairing. We analyze the resulting 
$^1S_0$ pairing gap in isospin-symmetric nuclear matter and apply 
the RHB model to the calculation of ground-state pairing properties of 
finite spherical nuclei. 

In Sec.~\ref{SecRHB} we present an outline of the RHB model and
introduce the renormalization scheme for the case of zero-range pairing.
The model is applied in Sec.~\ref{SecNM} to pairing in 
isospin-symmetric nuclear matter. Ground-state 
pairing properties of Sn nuclei are analyzed in Sec.~\ref{SecSn}. 
Sec.~\ref{SecSum} contains the summary and conclusions.

\section{\label{SecRHB} Relativistic Hartree-Bogoliubov model with 
zero-range pairing}

A detailed review of the relativistic Hartree-Bogoliubov model 
can be found, for instance, in Ref.~\cite{VALR.05}. 
In this section we include those features which are essential 
for the discussion of the renormalization of the RHB equations.
The model can be derived within the framework of
covariant density functional theory. When pairing correlations are
included, the energy functional depends not only on the density
matrix $\hat{\rho}$ and the meson fields $\phi_{m}$, but in addition
also on the anomalous density $\hat{\kappa}$
\begin{equation}
E_{RHB}[\hat{\rho},\hat{\kappa},\phi_{m}]=E_{RMF}[\hat{\rho},\phi
_{m}]+E_{pair}[\hat{\kappa}]\;,\label{ERHB}%
\end{equation}
where $E_{RMF}[\hat{\rho},\phi]$ is the RMF energy density functional 
and the pairing energy $E_{pair}[\hat{\kappa}]$
is given by
\begin{equation}
E_{pair}[\hat{\kappa}]=\frac{1}{4}\mathrm{Tr}\left[  \hat{\kappa}^{\ast}%
V^{pp}\hat{\kappa}\right]  .
\end{equation}
$V^{pp}$ denotes a general two-body pairing interaction. The
equation of motion for the generalized density matrix 
\begin{equation}
\mathcal{R}=\left(
\begin{array}
[c]{cc}%
\rho & \kappa\\
-\kappa^{\ast} & 1-\rho^{\ast}
\end{array}
\right)  \;,
\end{equation}
reads
\begin{equation}
i\partial_{t}\mathcal{R}=\left[
\mathcal{H}(\mathcal{R}),\mathcal{R}\right]
\;.\label{TDRHB}
\end{equation}
The generalized Hamiltonian $\mathcal{H}$ is a functional derivative of the 
energy with respect to the generalized density
\begin{equation}
\mathcal{H}_{RHB}~=~\frac{\delta
E_{RHB}}{\delta\mathcal{R}}~=~\left(
\begin{array}
[c]{cc}%
\hat{h}_{D}-m-\mu & \hat{\Delta}\\
-\hat{\Delta}^{\ast} & -\hat{h}^{\ast}_{D}+m+\mu
\end{array}
\right)  \;.\label{HFB-hamiltonian}%
\end{equation}
The self-consistent mean field $\hat{h}_{D}$ is the Dirac
Hamiltonian, and the pairing field reads
\begin{equation}
\Delta_{ab}(\mathbf{r},\mathbf{r}^{\prime})={\frac{1}{2}}\sum\limits_{c,d}%
V_{abcd}^{pp}(\mathbf{r},\mathbf{r}^{\prime})\kappa_{cd}(\mathbf{r}%
,\mathbf{r}^{\prime}),\label{equ.2.5}%
\end{equation}
where $a,b,c,d$ denote quantum numbers that specify the Dirac
indices of the spinors, and
$V_{abcd}^{pp}(\mathbf{r},\mathbf{r}^{\prime})$ are the matrix
elements of a general two-body pairing interaction.

Pairing effects in nuclei are restricted to an energy window of a few MeV 
around the Fermi level, and their scale
is well separated from the scale of binding energies, which
are in the range of several hundred to thousand MeV. 
There is no experimental evidence for any
relativistic effect in the nuclear pairing field $\hat{\Delta}$.
Therefore, pairing can be treated as a non-relativistic phenomenon, and a 
hybrid RHB model with a non-relativistic pairing interaction can be employed.
For a general two-body interaction, the matrix elements of 
the relativistic pairing field read
\begin{equation}
\hat{\Delta}_{a_1 p_1, a_2 p_2} =  
{\frac{1}{2}}\sum\limits_{a_3 p_3, a_4 p_4}
\langle a_1 p_1, a_2 p_2 |V^{pp}|a_3 p_3, a_4 p_4\rangle_a~
\kappa_{a_3 p_3, a_4 p_4}\; ,
\end{equation}
where the indices ($p_1,p_2,p_3,p_4 = +,-$) refer to the large and 
small components of the quasiparticle Dirac spinors. In most 
applications of the RHB model, only the large components of the 
spinors $U_{k}({\bf r})$ and $V_{k}({\bf r})$ have been included in 
the non-relativistic pairing tensor $\hat{\kappa}$ in Eq.
(\ref{kappa0}). The resulting pairing 
field reads
\begin{equation}
\hat{\Delta}_{a_1 +, a_2 +} =  
{\frac{1}{2}}\sum\limits_{a_3 +, a_4 +}
\langle a_1 +, a_2 + |V^{pp}|a_3 +, a_4 +\rangle_a~
\kappa_{a_3 +, a_4 +}\; .
\end{equation}
The other components: $\hat{\Delta}_{+-}$,
$\hat{\Delta}_{-+}$, and $\hat{\Delta}_{--}$ are neglected, 
in accordance with the results that are obtained with 
a relativistic zero-range force \cite{SR.02}.  

The ground state of an open-shell nucleus is described by the solution of
the relativistic Hartree-Bogoliubov equations
\begin{equation}
\left(
\begin{array}
[c]{cc}%
\hat{h}_{D}-m-\mu & \hat{\Delta}\\
-\hat{\Delta}^{\ast} & -\hat{h}^{\ast}_{D}+m+\mu
\end{array}
\right)  \left(
\begin{array}
[c]{c}%
U_{k}(\mathbf{r})\\
V_{k}(\mathbf{r})
\end{array}
\right)  =E_{k}\left(
\begin{array}
[c]{c}%
U_{k}(\mathbf{r})\\
V_{k}(\mathbf{r})
\end{array}
\right)  \;, \label{eqhb}%
\bigskip
\end{equation}
which correspond to the stationary limit of Eq. (\ref{TDRHB}).

The chemical potential $\mu$ is determined by the particle
number subsidiary condition in order that the expectation value of
the particle number operator in the ground state equals the number
of nucleons. The column vectors denote the quasiparticle wave
functions, and $E_{k}$ are the quasiparticle energies. 
The RHB wave functions determine the hermitian single-particle density matrix
\begin{equation}
\hat{\rho}_{ll^{\prime}}=(V^{\ast}V^{T})_{ll^{\prime}},%
\label{rho0}%
\end{equation}
and the antisymmetric anomalous density
\begin{equation}
\hat{\kappa}_{ll^{\prime}}=(V^{\ast}U^{T})_{ll^{\prime}}. \label{kappa0}%
\end{equation}

The calculated nuclear ground-state properties sensitively depend 
on the choice 
of the effective Lagrangian and pairing interaction.
Over the years many parameter sets of the
mean-field Lagrangian have been derived that provide a satisfactory
description of nuclear properties along the $\beta $-stability line.
The most successful RMF effective interactions are purely
phenomenological, with parameters adjusted to reproduce the nuclear
matter equation of state and a set of global properties of spherical
closed-shell nuclei. This framework has recently been 
extended to include effective 
Lagrangians with explicit density-dependent meson-nucleon 
couplings. In a number of studies 
it has been shown that this class of  global
effective interactions provides an improved description of asymmetric 
nuclear matter, neutron matter and finite nuclei far from stability. 
In the present analysis of ground-state properties of Sn isotopes 
the density-dependent effective 
interaction DD-ME1~\cite{Nik1.02} will be employed in the
particle-hole ($ph$) channel of the RHB model.

In the following we extend the regularization scheme of Bulgac and Yu 
\cite{Bul1,Bul2,Bul3} to the solution of the relativistic 
Hartree-Bogoliubov equations for a zero-range
pairing interaction 
\begin{equation}
V^{pp}(\mathbf{r},\mathbf{r}^{\prime}) = g \delta(\mathbf{r} -
\mathbf{r}^{\prime})\; .
\label{pair_int}
\end{equation}
In Refs.~\cite{Bul1,Bul2} it has been shown that in this 
case the renormalized pairing field can be expressed as
\begin{equation}
\Delta (\mathbf{r}) = 
   -g_{\mathit{eff}}(\mathbf{r}) \kappa_c (\mathbf{r}) \;,
\label{eq:Delta}
\end{equation}
where $\kappa_c(\mathbf{r})$ denotes the cut-off anomalous density
\begin{equation}
\kappa_c(\mathbf{r})    =\sum\limits_{E_k>0}^{E_c} 
V_{k}^{\dagger}(\mathbf{r})U_{k}(\mathbf{r}) \;.
\label{pair_c}
\end{equation}
The cut-off energy $E_c$ defines the two corresponding momenta $k_c$ and $l_c$
\begin{eqnarray}
\sqrt{k_c^2(\mathbf{r}) + {m^*}^2(\mathbf{r})}+ V(\mathbf{r}) - m &=&E_c +\mu
\label{cut1} ,\\
\sqrt{l_c^2(\mathbf{r}) + {m^*}^2(\mathbf{r})}+ V(\mathbf{r}) - m &=&-E_c+\mu 
\label{cut2}\;.
\end{eqnarray}
$m^*(\mathbf{r}) = m +S(\mathbf{r})$ is the Dirac mass, and $S(\mathbf{r})$
and $V(\mathbf{r})$ are, respectively, 
the scalar and vector single-nucleon potentials contained 
in the Dirac Hamiltonian $\hat{h}_{D}$. The chemical potential $\mu$ determines
the local Fermi momentum 
\begin{equation}
\sqrt{k_f^2(\mathbf{r}) + {m^*}^2(\mathbf{r})} + V(\mathbf{r}) - m =\mu\;. 
\label{kfermi}
\end{equation}
The effective, position-dependent coupling in Eq.~(\ref{eq:Delta}) reads
\begin{equation}
\frac{1}{ g_{\mathit{eff}}(\mathbf{r})} =
\frac{1}{g} + F_1(\mathbf{r}) + F_2(\mathbf{r}) \;,
\label{g_eff} 
\end{equation}
with
\begin{eqnarray}
F_1(\mathbf{r}) &=& -{{k_c(\mathbf{r})\sqrt{k_f^2(\mathbf{r})+
{m^*}^2(\mathbf{r})}}\over{4\pi^2}} \left [ 1 - 
{{k_f(\mathbf{r})}\over{k_c(\mathbf{r})}}~{\rm Ar~cth} 
{{k_c(\mathbf{r})}\over{k_f(\mathbf{r})}} \right ] \nonumber \\
&&-{{k_c(\mathbf{r})\sqrt{k_c^2(\mathbf{r})+
{m^*}^2(\mathbf{r})}}\over{8\pi^2}} +
{{\sqrt{k_f^2(\mathbf{r})+
{m^*}^2(\mathbf{r})}}\over{4\pi^2}}~k_f(\mathbf{r}) ~{\rm Ar~cth}
{{k_c(\mathbf{r})\sqrt{k_f^2(\mathbf{r})+
{m^*}^2(\mathbf{r})}}\over {k_f(\mathbf{r})\sqrt{k_c^2(\mathbf{r})+
{m^*}^2(\mathbf{r})}}} \nonumber \\
&&-{{2k_f^2(\mathbf{r})+
{m^*}^2(\mathbf{r})}\over {8\pi^2}} ~{\rm ln} 
{{k_c(\mathbf{r})+ \sqrt{k_c^2(\mathbf{r})+
{m^*}^2(\mathbf{r})}}\over {{m^*}(\mathbf{r})}} 
\label{F1Re}
\end{eqnarray}
\begin{eqnarray}
F_2(\mathbf{r}) &=& -{{l_c(\mathbf{r})\sqrt{k_f^2(\mathbf{r})+
{m^*}^2(\mathbf{r})}}\over{4\pi^2}} \left [ 1 - 
{{k_f(\mathbf{r})}\over{l_c(\mathbf{r})}}~{\rm Ar~th} 
{{l_c(\mathbf{r})}\over{k_f(\mathbf{r})}} \right ] \nonumber \\
&&-{{l_c(\mathbf{r})\sqrt{l_c^2(\mathbf{r})+
{m^*}^2(\mathbf{r})}}\over{8\pi^2}} +
{{\sqrt{k_f^2(\mathbf{r})+
{m^*}^2(\mathbf{r})}}\over{4\pi^2}}~k_f(\mathbf{r}) ~{\rm Ar~th}
{{l_c(\mathbf{r})\sqrt{k_f^2(\mathbf{r})+
{m^*}^2(\mathbf{r})}}\over {k_f(\mathbf{r})\sqrt{l_c^2(\mathbf{r})+
{m^*}^2(\mathbf{r})}}} \nonumber \\
&&-{{2k_f^2(\mathbf{r})+
{m^*}^2(\mathbf{r})}\over {8\pi^2}} ~{\rm ln} 
{{l_c(\mathbf{r})+ \sqrt{l_c^2(\mathbf{r})+
{m^*}^2(\mathbf{r})}}\over {{m^*}(\mathbf{r})}} 
\label{F2Re}
\end{eqnarray}
$F_1 + F_2$ is the relativistic generalization of the corresponding 
correction to the coupling constant $g$, as defined in 
Eq. (16) of Ref.~\cite{Bul1}.   
\section{\label{SecNM}Pairing properties of symmetric nuclear matter}

\begin{figure}
\begin{center}
\includegraphics[scale= 0.6,clip]{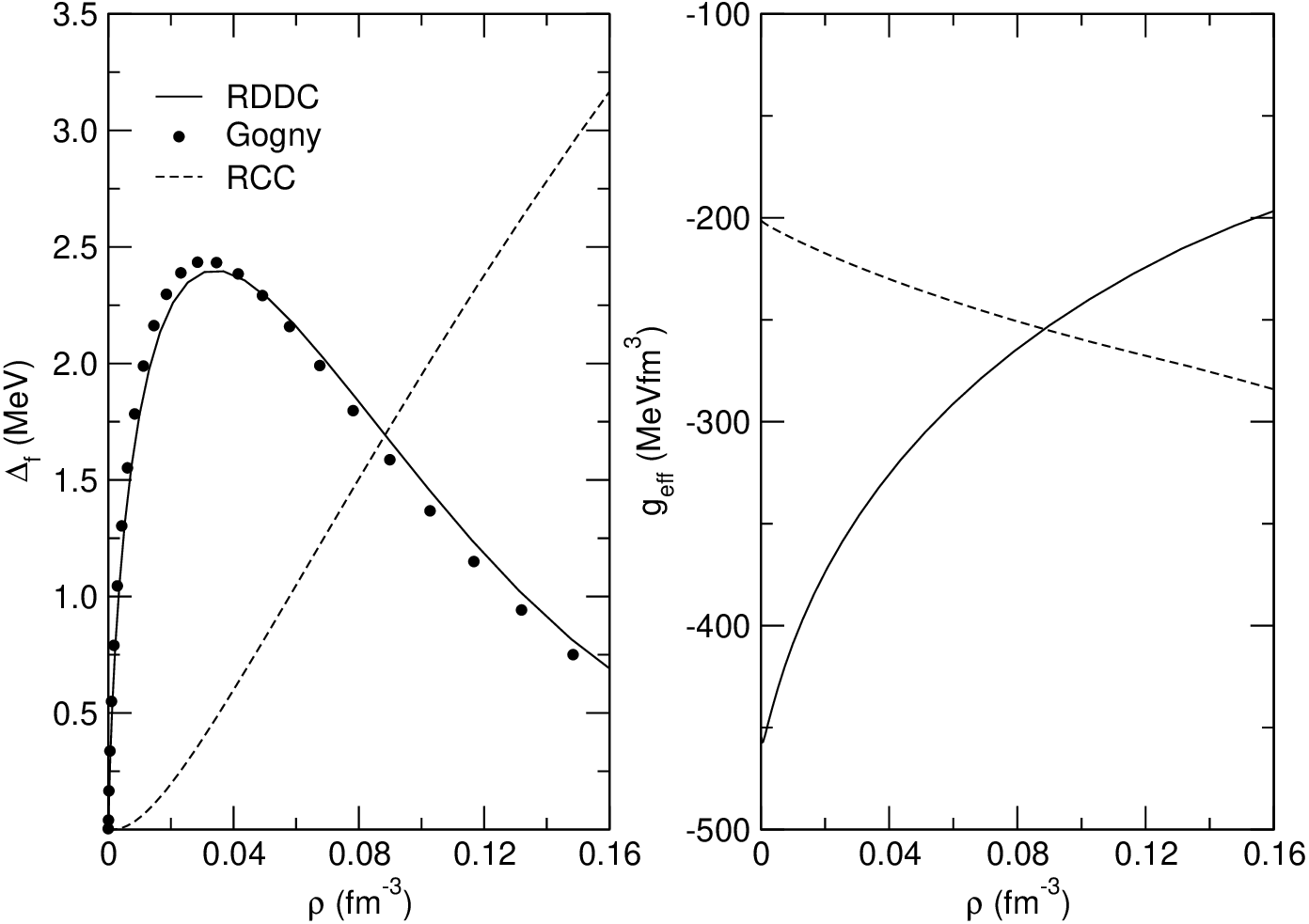}
\end{center}
\caption{Pairing gap in symmetric nuclear matter as a function of 
density for the zero-range interaction with constant 
coupling $g = -330$ MeV fm$^{3}$
(dashed) and the density-dependent coupling Eq.~(\protect\ref{g_rho}) 
(solid). The corresponding density-dependent curves $g_{\mathit{eff}}$ 
(Eq.~(\protect\ref{g_eff})) are plotted in the panel on the right.
The dots in the left panel denote the pairing gap calculated with 
the Gogny D1S interaction.}
\label{FigA}
\end{figure}

A zero-range pairing interaction leads to a particularly simple
expression for the gap equation in symmetric nuclear matter
\begin{equation}
\frac{1}{ g_{\mathit{eff}}} = 
- \frac{1}{4\pi^2}\int_{l_c}^{k_c} dk~ 
{k^2\over \sqrt{\left [ \sqrt{k^2+{m^*}^2} -
\sqrt{k_f^2+{m^*}^2}~ \right ]^2 + \Delta^2}} \; ,
\label{gap}
\end{equation}
The momenta $k_c$ and $l_c$ are determined by the cut-off energy $E_c$
Eqs. (\ref{cut1},~\ref{cut2}), and the effective coupling $g_{\mathit{eff}}$ 
is defined in Eq.~(\ref{g_eff}).
In the left panel of Fig.~\ref{FigA} we display the density dependence 
of the resulting pairing gap in nuclear matter (dashed curve).
The single-particle spectrum has been calculated with the
relativistic effective interaction DD-ME1~\cite{Nik1.02}, and 
the coupling constant of the zero-range pairing 
interaction Eq.~(\ref{pair_int}) $g = -330$ MeV fm$^{3}$ is 
typical for the values used by Bulgac and Yu in their 
analyses. The pairing gap is shown in comparison    
to the gap calculated with the effective Gogny
interaction D1S \cite{BGG.91} (dots). The corresponding 
single-particle spectrum has been computed in the
Hartree-Fock approximation for the Gogny interaction. The 
density dependence of the two gaps is completely different. 
The pairing gap of the renormalized zero-range interaction 
increases uniformly with density, whereas the gap of the 
Gogny interaction display the characteristic maximum of 
$\approx 2.5$ MeV at low density $\rho = 0.03 - 0.04$ fm$^{-3}$
(corresponding to a Fermi momentum of approximately 0.8 fm$^{-1}$) 
and decreases at higher densities. The bell-shaped form of the pairing gap
as a function of the density was, in fact, 
obtained already more than forty years ago \cite{ES.60}.
This density dependence
is not characteristic only of the phenomenological 
finite-range interactions, 
but is also obtained when the gap is calculated with 
bare nucleon-nucleon potentials adjusted to 
the empirical nucleon-nucleon phase shifts and 
deuteron properties
(for a recent review see Ref.~\cite{DH-J.03}).
The decrease of the gap at Fermi momenta $k_f > 0.8$ fm$^{-1}$
simply reflects the repulsive character of the nucleon-nucleon 
interaction at short distances \cite{SRR.02}. 
Of course there is no repulsive component in the zero-range 
force with constant coupling Eq.~(\ref{pair_int}), and 
the corresponding pairing gap displays the unphysical 
uniform increase with density. We notice, however, that 
in the range of densities shown in Fig.~\ref{FigA}, i.e. up 
to nuclear matter saturation density, the values of the pairing
gap of the renormalized zero-range interaction are comparable 
with those of the Gogny pairing gap. As will be shown in the 
next section, this means that the renormalization scheme 
for the zero-range interaction with constant coupling can be 
safely applied to the calculation of pairing correlations in 
finite nuclei, provided an appropriate choice is made 
for the strength parameter $g$.

On the other hand, there is no particular reason why the 
strength parameter $g$ of the zero-range pairing interaction should 
be a constant. In fact, in many applications to finite nuclei 
an explicit density dependence is introduced, and in this 
way pairing correlations partially include finite-range effects. For instance, 
in one of the first applications \cite{BE.91} Bertsch and Esbensen 
used a density-dependent contact interaction, together with 
a simple energy cut-off, in a description of pairing correlations in 
weakly bound neutron-rich nuclei. They also compared the corresponding
pairing gap in symmetric nuclear matter with the result of a Hartree-Fock
calculation using the Gogny interaction. In the present anaysis we have 
adjusted a density-dependent strength parameter $g(\rho)$ 
of the zero-range pairing interaction Eq.~(\ref{pair_int}), 
in such a way that the pairing gap of the renormalized 
zero-range interaction Eq.~(\ref{gap}), reproduces the density
dependence of the Gogny pairing gap. The resulting density dependence 
can be approximated by the following analytic expression
\begin{equation}
g(\rho) = \frac{1}{a_0 + a_1 \rho^{1/3} + a_2 \rho^{2/3}} \; ,
\label{g_rho}
\end{equation}
with $a_0 = -0.064$ fm$^{-2}$, $a_1 = 0.447$ fm$^{-1}$, and $a_2 =-3.693$. 
The resulting pairing gap, displayed in the left panel of 
Fig.~\ref{FigA} (solid line),
is in very good agreement with the one calculated using the Gogny interaction.
A very similar procedure was employed in Ref.\cite{YuB.03}, where the density 
dependence of the ``bare coupling constant" $g(\rho)$ was adjusted to a 
specific formula for the pairing gap in low-density homogeneous neutron matter.

In the right panel of Fig.~\ref{FigA} we display the effective couplings
$g_{\mathit{eff}}$ calculated using the constant  
$g = -330$ MeV fm$^{3}$, and the density dependent coupling 
of Eq. (\ref{g_rho}). The density dependence of the two 
effective couplings is completely different.
In order to prevent an unphysical growth of 
the pairing gap with density, the density dependence of the pairing 
strength Eq. (\ref{g_rho}) ensures that the effective coupling 
becomes weaker with increasing nucleon density. A very strong 
effective coupling in the low-density region produces a peak 
in the corresponding pairing gap shown in the left panel. 
On the other hand, $g_{\mathit{eff}}$ calculated using the constant
coupling increases in absolute value with density, i.e. the 
resulting pairing gap increases uniformly with density. However,
rather similar values for the two effective couplings 
$g_{\mathit{eff}}$ are calculated
in the region of densities characteristic for the bulk of finite nuclei.
One should not, therefore, expect very different results for 
the pairing properties of finite nuclei calculated with the 
zero-range interaction with constant coupling, or with the 
density-dependent coupling of Eq. (\ref{g_rho}). In the next 
section we will show that this is really not true in 
weakly-bound nuclei far from stability.

\begin{figure}
\begin{center}
\includegraphics[scale= 0.6,clip]{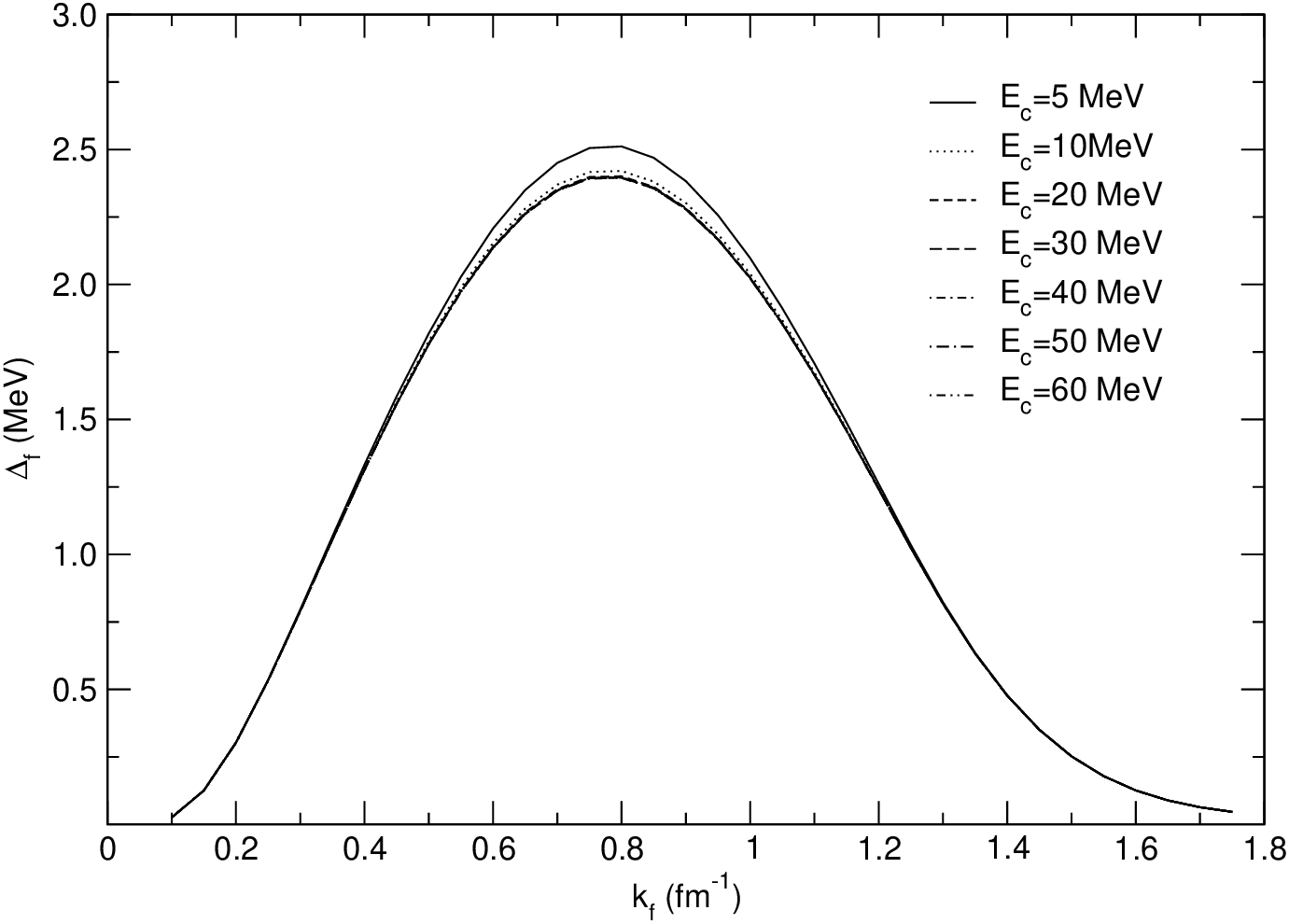}
\end{center}
\caption{Pairing gap in symmetric nuclear matter as a function of 
the Fermi momentum, calculated with the zero-range interaction 
and the density-dependent coupling Eq.~(\protect\ref{g_rho}), 
for a series of cut-off energies.}
\label{FigB}
\end{figure}

The renormalization prescription must, of course, lead to a pairing field
which is independent of the cut-off energy $E_c$, if the latter is 
chosen large enough. This is illustrated in Fig.~\ref{FigB}, 
where we plot the pairing gap,
calculated using the density-dependent coupling of Eq.~(\ref{g_rho}), 
for a number of
characteristic values $E_c$ in the interval between 5 MeV and 60 MeV.
The pairing gap shows a weak dependence on the cut-off energy only for the
two lowest values of $E_c$. When the cut-off is
increased beyond 10 MeV, the corresponding pairing gaps cannot
be distinguished. Thus already for $E_c \geq 10$ MeV the pairing 
gap of the renormalized zero-range interaction in symmetric 
nuclear matter converges. This is in agreement with the results
obtained in the analysis of the pairing gap in homogeneous neutron
matter~\cite{Bul1}.
 
\section{\label{SecSn}Ground-state pairing properties of spherical nuclei}

In this section the renormalization scheme is tested in the calculation
of ground-state pairing properties of Sn isotopes.
The DD-ME1 mean-field Lagrangian is employed for the $ph$ channel, and the 
zero-range interaction Eq.~(\ref{pair_int}) is used in the $pp$ channel.
The renormalization procedure described in the previous section is 
carried out for the zero-range interaction with constant pairing strength 
$g = -330$ MeV fm$^{3}$, and and for the density-dependent 
coupling of Eq.~(\ref{g_rho}). In the latter case
the density dependence of the pairing strength  has been
adjusted to reproduce the Gogny D1S pairing gap 
in symmetric nuclear matter. In the following we denote by 
RCC the case of the renormalized constant coupling, and by RDDC the
results obtained with the renormalized density-dependent coupling.

While in the symmetric nuclear matter the Fermi momentum is always real 
(see Eq. (\ref{kfermi})), in the surface region of finite nuclei it becomes 
imaginary. In Ref.~\cite{Bul1} it has been shown that 
also in this case the renormalized anomalous density is
real. The effective coupling $g_{\mathit{eff}}$ is still given by 
Eq.~(\ref{g_eff}), but 
\begin{eqnarray}
F_1(\mathbf{r}) &=& -{{k_c(\mathbf{r})\sqrt{-|k_f(\mathbf{r})|^2+
{m^*}^2(\mathbf{r})}}\over{4\pi^2}} \left [ 1 - 
{{k_f(\mathbf{r})}\over{k_c(\mathbf{r})}}~{\rm Ar~ctg} 
{{k_c(\mathbf{r})}\over{k_f(\mathbf{r})}} \right ] \nonumber \\
&&-{{k_c(\mathbf{r})\sqrt{k_c^2(\mathbf{r})+
{m^*}^2(\mathbf{r})}}\over{8\pi^2}} +
{{\sqrt{-|k_f(\mathbf{r})|^2+
{m^*}^2(\mathbf{r})}}\over{4\pi^2}}~k_f(\mathbf{r}) ~{\rm Ar~ctg}
{{k_c(\mathbf{r})\sqrt{-|k_f(\mathbf{r})|^2+
{m^*}^2(\mathbf{r})}}\over {k_f(\mathbf{r})\sqrt{k_c^2(\mathbf{r})+
{m^*}^2(\mathbf{r})}}} \nonumber \\
&&-{{(-2|k_f(\mathbf{r})|^2+
{m^*}^2(\mathbf{r}))}\over {8\pi^2}} ~{\rm ln} 
{{k_c(\mathbf{r})+ \sqrt{k_c^2(\mathbf{r})+
{m^*}^2(\mathbf{r})}}\over {{m^*}(\mathbf{r})}} \;, 
\label{F1Im}
\end{eqnarray}
and 
\begin{equation}
F_2(\mathbf{r}) = 0\;.
\label{F2Im}
\end{equation}
However, if either $k_c$ or $l_c$ becomes imaginary,
the corresponding terms in the effective coupling should be omitted.

\begin{figure}
\begin{center}
\includegraphics[scale= 0.6,clip]{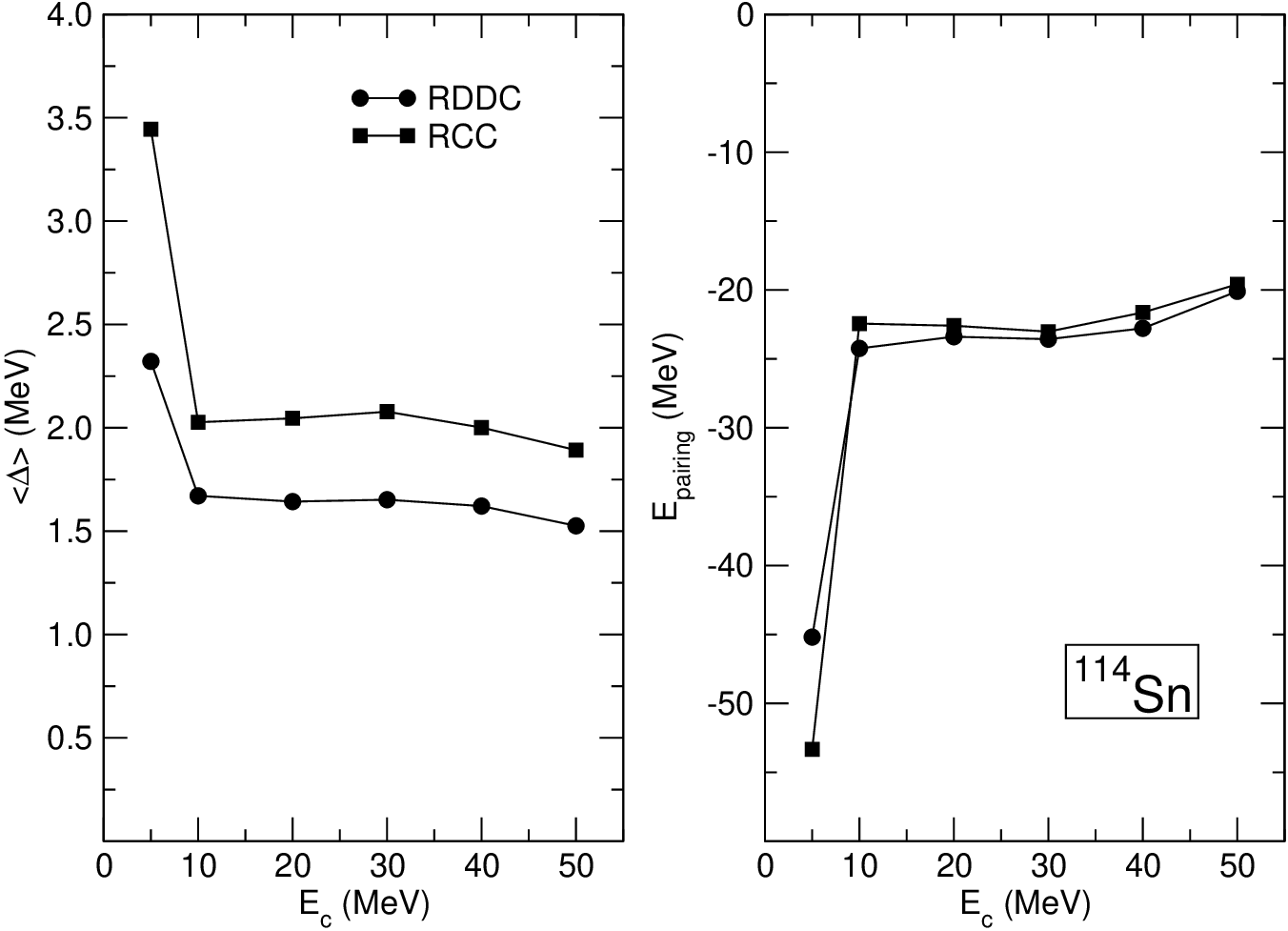}
\end{center}
\caption{Average neutron pairing gaps (left), and pairing energies (right) 
for $^{114}$Sn, calculated with a zero-range interaction, as functions 
of the cut-off energy. The calculations are performed  with the renormalization 
of the constant coupling $g = -330$ MeV fm$^{3}$ (squares), and with the 
renormalization of the density-dependent coupling  
Eq.~(\protect\ref{g_rho}) (dots).}
\label{FigC}
\end{figure}

The rate of convergence of the renormalization scheme is illustrated 
in Fig.~\ref{FigC} where, for the nucleus $^{114}$Sn, we display the
average pairing gaps and the pairing energies as functions
of the cut-off energy $E_c$. The average gaps shown in the
left panel, are defined as 
\begin{equation}
< \Delta_N > = {{\sum_{nlj} \Delta_{nlj} v_{nlj}^2}\over
		  {\sum_{nlj}  v_{nlj}^2}} \; ,
\label{ang}
\end{equation}
where $v_{nlj}^2$ are the occupation probabilities 
of the neutron states in the canonical basis. Both the 
pairing gaps and the pairing energies converge already for $E_c \geq 10$ MeV. 
We also notice that, even though the renormalized constant coupling 
and the renormalized density-dependent coupling lead to very 
different pairing gaps in symmetric nuclear matter, in $^{114}$Sn
they produce similar average pairing gaps and virtually identical 
pairing energies. 

\begin{figure}
\begin{center}
\includegraphics[scale= 0.6,clip]{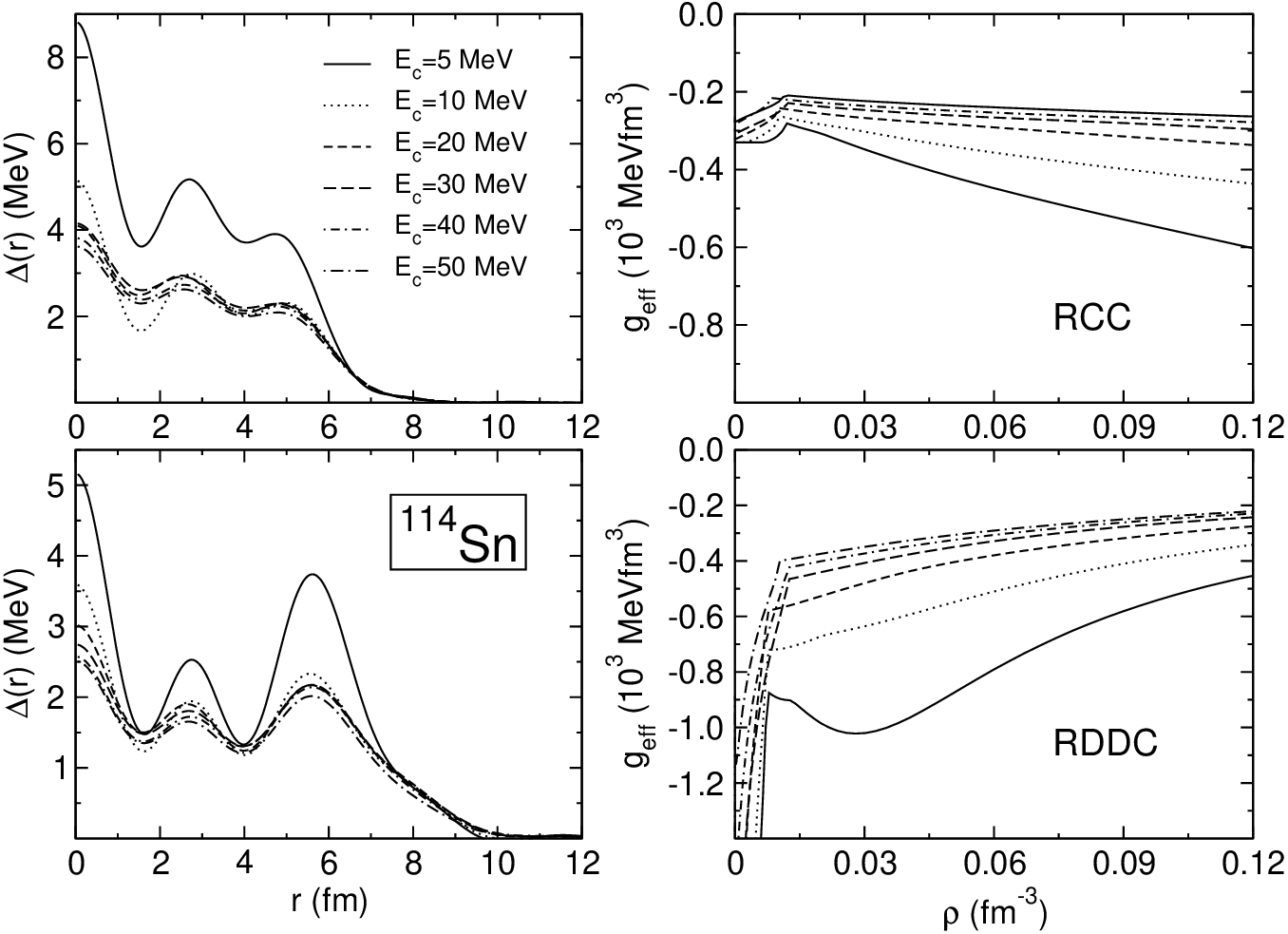}
\end{center}
\caption{The pairing fields as functions of the radial 
coordinate (left), and the curves $g_{\mathit{eff}}(r(\rho))$ 
(Eq.~(\protect\ref{g_eff})) as functions of the 
corresponding density in $^{114}$Sn (right), for a series of energy cut-offs. 
The upper panels display results of the renormalization procedure for 
the zero-range force with constant coupling $g = -330$ MeV fm$^{3}$
(RCC), and the lower ones correspond to the 
density-dependent coupling Eq.~(\protect\ref{g_rho}) (RDDC).}
\label{FigD}
\end{figure}

\begin{figure}
\begin{center}
\includegraphics[scale= 0.6,clip]{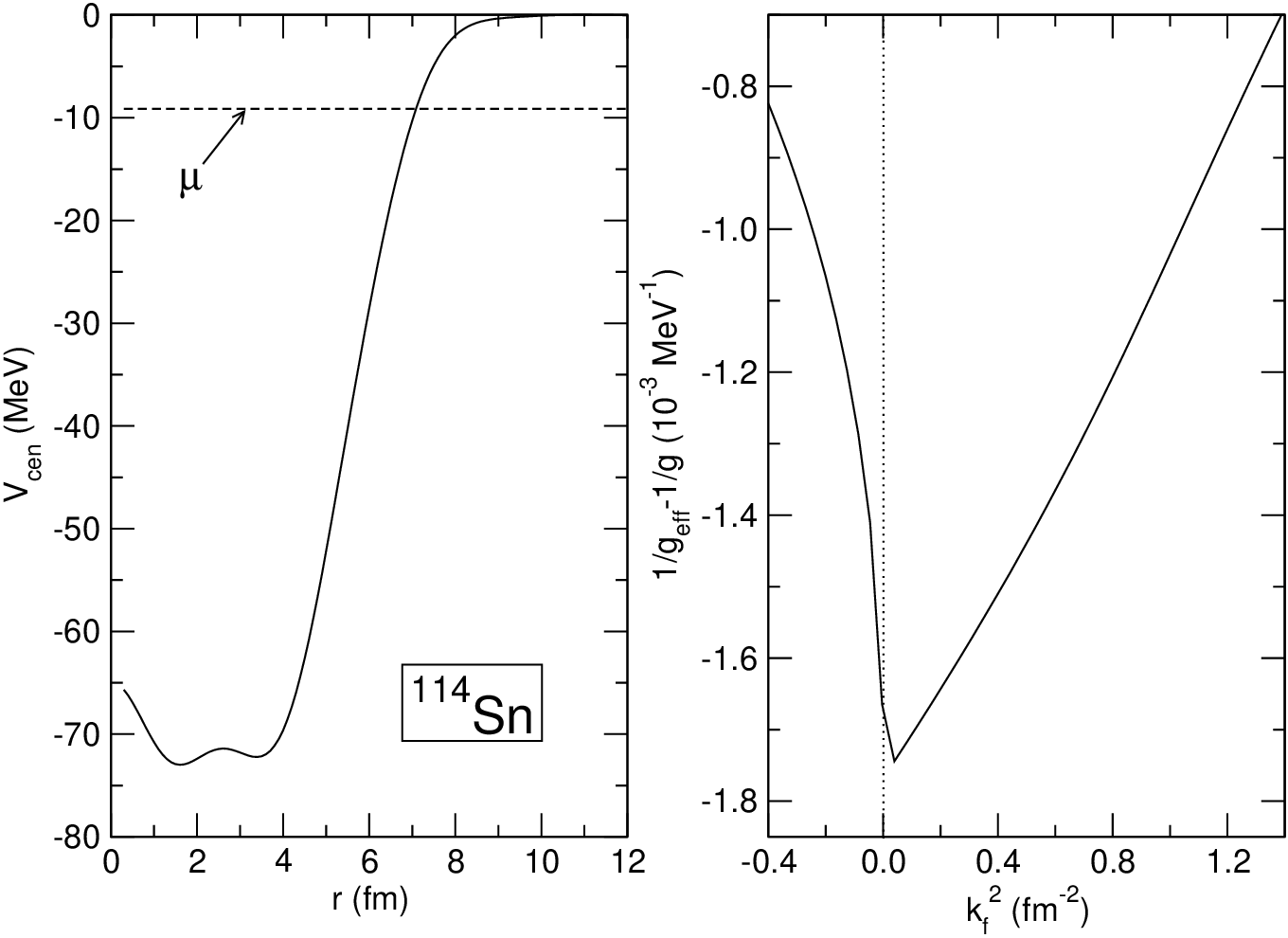}
\end{center}
\caption{The effective single-nucleon potential in $^{114}$Sn as a function of
the radial coordinate (left), and the correction to the 
strength parameter of the zero-range effective force (right) as 
a function of the square of the Fermi momentum.
$\mu$ denotes the position of the chemical potential.}
\label{FigE}
\end{figure}

The corresponding pairing fields as functions of the radial 
coordinate, and $g_{\mathit{eff}}$ Eq.~(\ref{g_eff}) as functions 
of the density, are plotted in Fig.~\ref{FigD} for a series of values
of the energy cut-off. In both cases the calculation of the pairing 
field and $g_{\mathit{eff}}$ shows convergence for $E_c > 10$ MeV. 
While the renormalized constant coupling and the renormalized 
density-dependent coupling produce very similar average pairing gaps 
and pairing energies, the dependence of the corresponding 
pairing fields on the radial coordinate is rather different.  
The RCC pairing field (upper left panel) is concentrated in the 
bulk of the nucleus, whereas the RDDC pairing field (lower left panel) 
exhibits a pronounced peak on the surface. This behavior
reflects the difference between the effective couplings $g_{\mathit{eff}}$,
already shown in the right panel of Fig.~\ref{FigA} for the case of 
symmetric nuclear matter. In the panels on the right of Fig.~\ref{FigD} we 
plot the effective couplings $g_{\mathit{eff}}(r(\rho))$ 
as functions of the nucleon
density in $^{114}$Sn. The $g_{\mathit{eff}}$ which corresponds to
the density-dependent coupling of Eq.~(\ref{g_rho}) 
decreases steeply in the region of very low density, i.e., 
on the surface of the nucleus. Consequently, also the pairing field 
displays a peak in the surface region.
In both the RCC and RDDC cases the pronounced discontinuity of 
the effective coupling $g_{\mathit{eff}}$ at very low density corresponds
to the transition from real to imaginary Fermi momentum $k_f$. 
This is illustrated in Fig.~\ref{FigE}, where we plot the 
effective single-nucleon
potential (left panel) and the correction to the coupling originating from
the renormalization of the anomalous density (right panel). The effective
single-nucleon potential is determined by the sum of the vector and 
scalar potentials $V_{cen}(\mathbf{r})=S(\mathbf{r})+V(\mathbf{r})$. 
For real values of the Fermi momentum (the effective potential is below the
chemical potential $\mu$) the correction to the coupling 
$F_1(\mathbf{r})+F_2(\mathbf{r})$ is calculated from Eqs. (\ref{F1Re})
and (\ref{F2Re}), and for imaginary values of the Fermi momentum
(the effective potential is above the chemical potential $\mu$) from
Eqs. (\ref{F1Im}) and (\ref{F2Im}). 
In the region where the Fermi momentum changes from real to imaginary
the correction $F_1(\mathbf{r})+F_2(\mathbf{r})$ displays a very
sharp peak, which is reflected in the discontinuities of the 
effective couplings. 

\begin{figure}
\begin{center}
\includegraphics[scale= 0.6,clip]{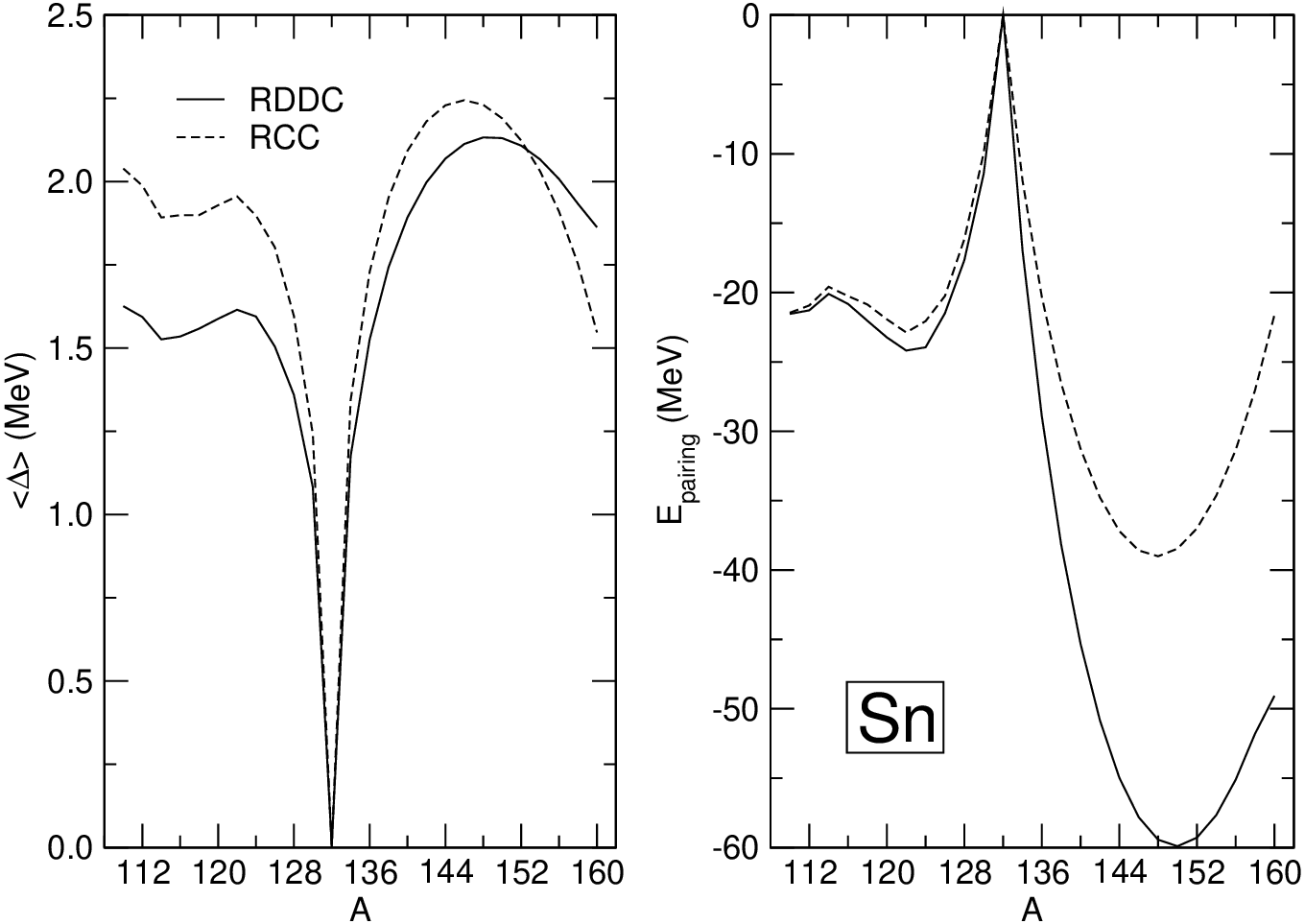}
\end{center}
\caption{Average neutron pairing gaps (left), and pairing energies (right)
for the chain of even-even Sn isotopes with $110 \le A \le 160$. 
The calculations are performed  with the renormalization 
of the constant coupling $g = -330$ MeV fm$^{3}$ (dashed), and with the 
renormalization of the density-dependent coupling  
Eq.~(\protect\ref{g_rho}) (solid).}
\label{FigF}
\end{figure}

\begin{figure}
\begin{center}
\includegraphics[scale= 0.6,clip]{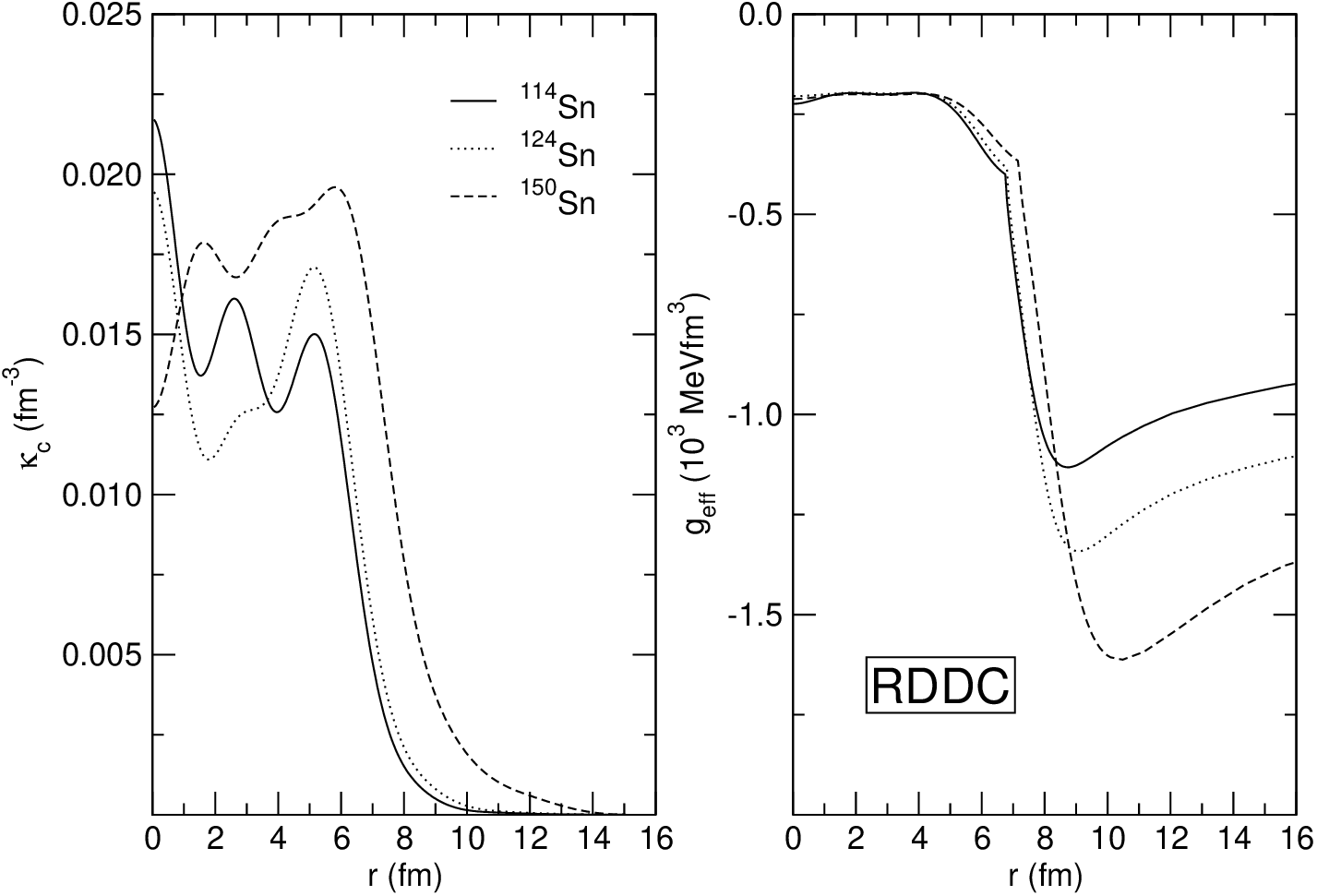}
\end{center}
\caption{The anomalous densities in $^{114}$Sn, $^{124}$Sn, $^{150}$Sn isotopes
as functions of the radial 
coordinate (left), and the density-dependent curves $g_{\mathit{eff}}$ 
(Eq.~(\protect\ref{g_eff})) (right). The calculations are 
performed  with the renormalization of the density-dependent coupling  
Eq.~(\protect\ref{g_rho}) (RDDC).}
\label{FigG}
\end{figure}

\begin{figure}
\begin{center}
\includegraphics[scale= 0.6,clip]{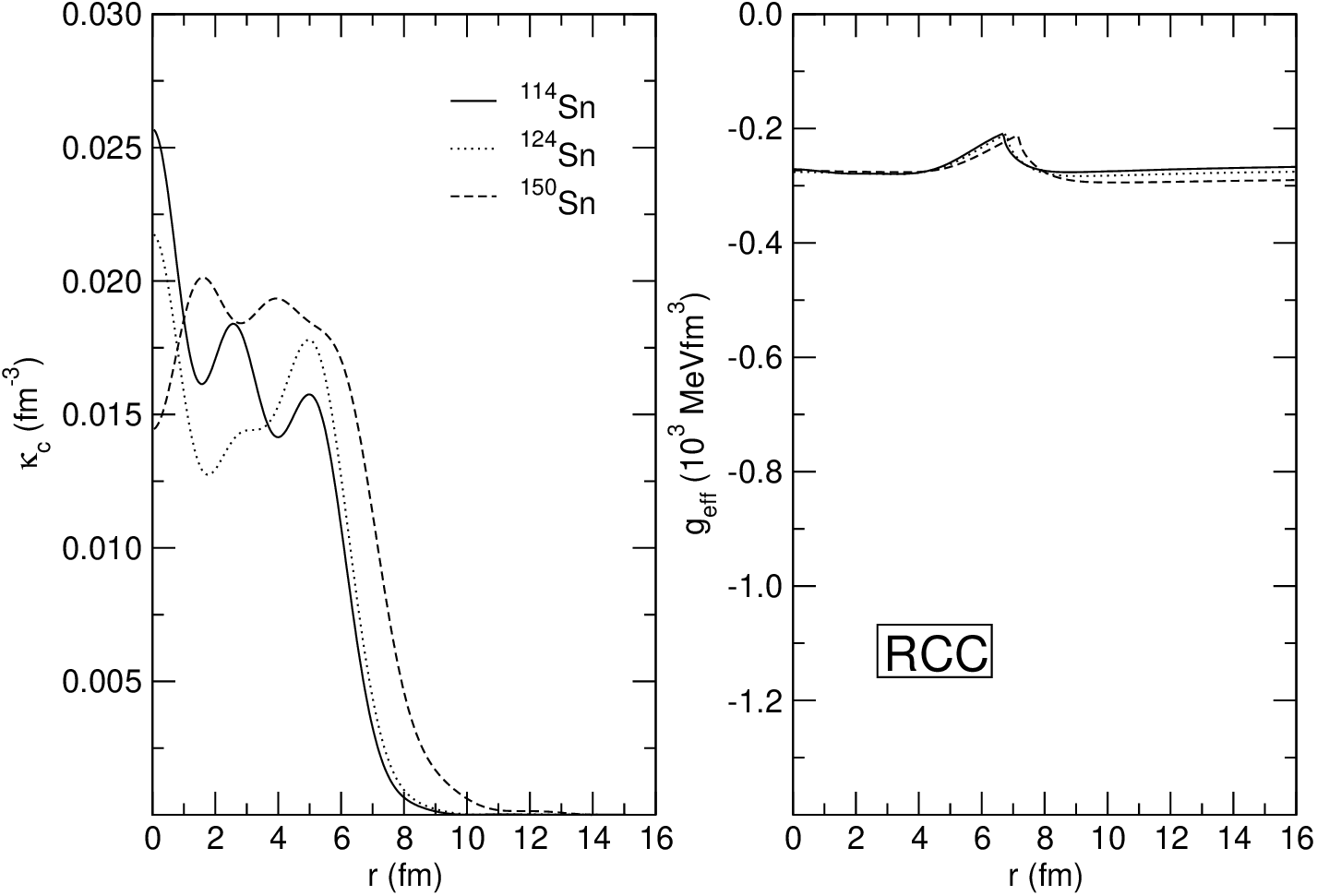}
\end{center}
\caption{Same as in Fig.~\protect\ref{FigG}, but for the 
RCC case with the constant coupling 
$g = -330$ MeV fm$^{3}$.}
\label{FigH}
\end{figure}

The importance of possible surface effects is illustrated in Fig.~\ref{FigF},
where we plot the calculated average pairing gaps and pairing energies
for the chain of even-even Sn isotopes with $110 \le A \le 160$. 
Although both the RCC and RDCC schemes lead to comparable values
of the average pairing gaps for the entire isotopic chain, the pairing energies
differ significantly for isotopes beyond the doubly closed-shell $^{132}$Sn. 
For example, the pairing energy of $^{150}$Sn calculated with renormalized 
density-dependent coupling (RDDC) is almost 25 MeV larger than the 
one calculated with the renormalized constant coupling (RCC). 
The large increase in the pairing energy for the RDDC case  
is caused by the dominant role of the surface region for the 
very neutron-rich Sn isotopes, and because the effective coupling 
is especially strong at very low densities.
In the panels on the left of Figs.~\ref{FigG} and~\ref{FigH} 
we plot the self-consistent solutions for the cut-off anomalous
densities Eq.~(\ref{pair_c}) for the isotopes 
$^{114}$Sn, $^{124}$Sn and $^{150}$Sn, calculated
using the RDDC and RCC effective couplings, respectively. 
The corresponding effective couplings $g_{\mathit{eff}}$
are shown in the panels on the right of Figs.~\ref{FigG} and~\ref{FigH}. 
The anomalous densities for $^{114}$Sn and $^{124}$Sn are concentrated in the
nuclear volume ($r \le 6$ fm), where the effective couplings
$g_{\mathit{eff}}$ have comparable values. Therefore, the corresponding pairing
energies are similar for the RDDC and RCC cases. In $^{150}$Sn,
on the other hand, the anomalous densities extend to the region 
$r\ge 8$ fm, where the RDDC effective 
coupling becomes much stronger than the one calculated with the 
constant coupling (RCC). Hence, the pairing energy for $^{150}$Sn,
calculated using the renormalized
density-dependent coupling is much larger than the one obtained with
the renormalized constant coupling.


\section{\label{SecSum}Conclusions}  
        
A simple renormalization scheme for the Hartree-Fock-Bogoliubov
equations with zero-range pairing has recently been introduced
\cite{Bul1,Bul2,Bul3}. In the present work we have implemented this 
renormalization scheme for the 
relativistic Hartree-Bogoliubov equations with a zero-range 
pairing interaction. The procedure is equivalent to a simple 
energy cut-off with a position dependent coupling constant.
We have verified that the resulting average pairing gaps and
pairing energies do not depend on the cut-off energy $E_c$, if
the latter is chosen large enough. Convergence is achieved already for 
values $E_c \ge 10$ MeV, both in nuclear matter and for finite nuclei.
If the strength parameter of the zero-range pairing is a constant,
the resulting pairing gap in symmetric nuclear matter displays an
unphysical increase with density. We have therefore
adjusted a density-dependent strength parameter of the zero-range pairing in 
such a way that the renormalization procedure reproduces 
in symmetric nuclear matter the pairing gap of the phenomenological
Gogny interaction. In this sense the present study goes beyond the 
simple extension of the renormalization scheme of Ref.~\cite{Bul1}
to the relativistic framework. However, the resulting effective coupling is 
too strong in the region of low density, and this leads to 
large pairing energies in open-shell nuclei with very diffuse surfaces, 
e.g. in neutron-rich Sn isotopes. One must therefore be careful 
when applying the renormalized HFB or RHB models with zero-range pairing 
to nuclei far from stability. Adjusting the strength parameter 
to the pairing gap in symmetric nuclear matter obviously 
does not provide enough information about the density dependence 
of the zero-range pairing to be used in very neutron-rich nuclei. 

\leftline{\bf ACKNOWLEDGMENTS}
This work has been supported in part by the Bundesministerium
f\"ur Bildung und Forschung - project 06 MT 193, by 
the Alexander von Humboldt Stiftung, 
and by the Croatian Ministry of Science - project 0119250.

\newpage

\bigskip

\begin{thebibliography}{999}
\bibitem{BHR.03} M. Bender, P.-H. Heenen, and P.-G. Reinhard,
	Rev. Mod. Phys. 75 (2003) 121.
	
\bibitem{VALR.05} D. Vretenar, A.V. Afanasjev, G.A. Lalazissis, and 
	P. Ring, Phys. Rep. (2005).
	
\bibitem{BGG.84} J.F. Berger, M. Girod, and D. Gogny, 
	Nucl. Phys. A 428,  23c (1984).
	
\bibitem{BGG.91} J. F. Berger, M. Girod, and D. Gogny, 
	Commput. Phys. Commun. 63, 365 (1991).
	
\bibitem{Meng.98} J. Meng, Nucl. Phys. A 635, 3 (1998).

\bibitem{Bul1} A. Bulgac, Phys. Rev. C 65, 051305 (2002).
	
\bibitem{Bul2} A. Bulgac and Y. Yu, Phys. Rev. Lett. 88, 042504 (2002).

\bibitem{Bul3} Y. Yu and A. Bulgac, Phys. Rev. Lett. 90, 222501 (2003).

\bibitem{Nik1.02} T. Nik\v si\' c, D. Vretenar, P. Finelli, and P. Ring,
	Phys. Rev. C 66, 024306 (2002).

\bibitem{SR.02} M. Serra and P. Ring, 
	Phys. Rev. {\bf C65},  064324  (2002).

\bibitem{SRR.02} M. Serra, A. Rummel, and P. Ring, 
	Phys. Rev. {\bf C65},  014304 (2002).
	
\bibitem{ES.60} V.J. Emery and A.M. Sessler, 
	Phys. Rev. {\bf 119},  248  (1960).
		   
\bibitem{DH-J.03} D.J. Dean and M. Hjorth-Jensen, 
	Rev. Mod. Phys. 75, 607 (2003); and references therein.
	
\bibitem{BE.91} G.F. Bertsch and H. Esbensen, 
	Ann. Phys. 209, 327 (1991).	
	
\bibitem{YuB.03} Y. Yu and A. Bulgac, 
	Phys. Rev. Lett. 90, 161101 (2003).	 
		 
		 
    	  	
\end{thebibliography}
\end{document}